\documentclass[fleqn,usenatbib]{mnras}

\usepackage{newtxtext,newtxmath}

\usepackage[T1]{fontenc}

\DeclareRobustCommand{\VAN}[3]{#2}
\let\VANthebibliography\thebibliography
\def\thebibliography{\DeclareRobustCommand{\VAN}[3]{##3}\VANthebibliography}

\usepackage{graphicx}	
\usepackage{amsmath}	

\usepackage{graphicx}
\usepackage{bm}
\usepackage{lipsum}
\usepackage{color, xcolor, ulem}	
\usepackage{longtable}
\usepackage{hyperref}

\title[$\delta$~Scuti stars in the Galactic bulge]{Modeling of multimode radially pulsating High-Amplitude $\delta$~Scuti stars from the OGLE Galactic bulge sample}

\author[H. Netzel, R. Smolec]
{H. Netzel$^{1,2,3}$\thanks{E-mail: henia@netzel.pl},
R. Smolec$^{3}$\\
$^{1}$Konkoly Observatory, Research Centre for Astronomy and Earth Sciences, E\"otv\"os Lor\'and Research Network (ELKH), MTA Centre of Excellence,\\ H-1121 Konkoly Thege Mikl\'os \'ut 15-17, Budapest, Hungary\\
$^{2}$MTA CSFK Lend\"ulet Near-Field Cosmology Research Group\\ H-1121 Konkoly Thege Mikl\'os \'ut 15-17, Budapest, Hungary\\
$^{3}$Nicolaus Copernicus Astronomical Centre, Polish Academy of Sciences, Bartycka 18, 00-716 Warsaw, Poland
}

\date{Accepted XXX. Received YYY; in original form ZZZ}

\pubyear{2022}

\begin{document}
\label{firstpage}
\pagerange{\pageref{firstpage}--\pageref{lastpage}}
\maketitle

\begin{abstract}
Thanks to relatively firm mode identification, possible based on period ratios only, High Amplitude Delta Scuti Stars pulsating in at least three radial modes are promising targets for asteroseismic inference. 
In this study we used the most numerous sample of HADS from the OGLE inner bulge fields that likely pulsate in either three or four radial modes simultaneously. We have computed a grid of pulsation models along evolutionary tracks and determined the physical parameters of stars by matching their pulsation periods and period ratios. For 176 HADS we determined physical parameters, i.e. masses, luminosities, effective temperatures, metallicities and ages. We present the distribution of physical parameters and discuss their properties. We selected 16 candidates for SX Phoenicis stars.

\end{abstract}




\section{Introduction}

$\delta$~Scuti stars are intermediate mass ($1.4-2.5\,{\rm  M}_\odot$) pulsating stars. They belong to Population I and have spectral types from A to early F. They are located on and close to the main sequence on the Hertzsprung-Russel diagram. Evolutionary status of the majority of them is core hydrogen burning, however, post-main-sequence evolution towards the red giant branch is also possible. Also stars at the pre-main-sequence (PMS) stage might show $\delta$~Scuti-type of pulsations. Objects like that are relatively rare among $\delta$ Scuti stars, however such PMS stars with $\delta$~Scuti-type of pulsations were already identified. First one was detected by \cite{breger1972} and search for these objects resulted in another discoveries \citep[e.g.][]{diaz_fraile2014} and focused studies of pre-main-sequence stars \citep[e.g.][]{zwintz2014}. On the other hand, $\delta$~Scuti stars evolving off the main sequence through the Hertzsprung gap are also observationally confirmed \citep{niu2021}.

Periods of pulsations of $\delta$~Scuti stars are in the 0.02--0.3d range. Pulsations of $\delta$~Scuti stars are mostly low-amplitude non-radial pressure modes, typically of low orders and low degrees. Frequency spectra are typically rich in excited modes, identification of which is problematic. 
Excitation mechanism is the same as in the rest of the classical instability strip, namely the $\kappa$ mechanism. Pulsation period of $\delta$~Scuti stars is usually a good criterion to distinguish them from pulsating stars located nearby on the Hertzsprung-Russel diagram: RR Lyrae, roAp or $\gamma$ Doradus. The instability strip for the latter class of pulsating stars overlaps with the instability strip for $\delta$~Scuti stars, which can result in so-called hybrid pulsators with pressure modes corresponding to the $\delta$~Scuti-type of pulsations and gravity modes corresponding the $\gamma$~Doradus-type of pulsations. Existence of such objects was confirmed by e.g. \cite{handler2009}.

A fraction of $\delta$~Scuti stars pulsate in fewer high-amplitude radial modes. These are called High-Amplitude Delta Scuti stars (HADS). Pulsations in HADS can be single-periodic with radial fundamental mode or the first overtone mode, but also double or triple periodic. Stars pulsating in four radial modes are also known \citep{pietrukowicz2013, netzel_hads}. Additionally, in the frequency spectra of HADS non-radial modes may also be present.
The incidence rate of HADS is estimated by \cite{lee2008} to be 0.24 per cent among Galactic $\delta$~Scuti stars.

Another interesting group overlapping with $\delta$~Scuti stars on the Hertzsprung-Russel diagram are SX Phoenicis stars (SX~Phe), which are considered typically to be Population II counterparts of $\delta$~Scuti stars \citep[see][and references therein]{catelan.smith2015}. SX~Phe often share pulsating properties with HADS. They pulsate in one or a few radial modes. They have typically smaller masses, around 1\,M$_\odot$, as shown in various modeling efforts \citep[e.g.][]{olech2005,pietrukowicz2013,antoci2019}.

The mass range of $\delta$~Scuti stars covers the interesting transition from main-sequence stars with radiative cores to stars with convective cores, hence giving a chance to study this transition in detail. Diversity of $\delta$~Scuti stars in evolutionary phases and in internal structure makes them interesting target for observations and asteroseismic modeling. However, only firmly identified pulsation modes may be used for asteroseismic inference. Identifying numerous non-radial modes present in $\delta$~Scuti stars is challenging and often prevents asteroseismic modeling of large samples. The exceptions are $\delta$~Scuti stars for which dedicated multiband and spectroscopic observations were carried out which enabled mode identification \citep[see e.g.][]{lenz2008}. Another exception is a group of $\delta$~Scuti stars that show regular patterns of frequencies giving chance for a mode identification \citep{bedding2020}. 
HADS give another chance to identify pulsation modes, due to the fact that radial modes have specific period ratios and can be identified through grouping in the Petersen diagram (diagram of shorter-to-longer period ratio versus the logarithm of the longer period). Multimode radial pulsations of HADS enabled \cite{pietrukowicz2013} to determine basic physical parameters of seven such stars.	

Only multi-mode pulsations are useful for asteroseismic study. Since the incidence rate of HADS is low, and the incidence rate of multimode HADS is even lower, the best place to look for these stars are large-scale photometric surveys. \cite{pietrukowicz2020} used Optical Gravitational Lensing Experiment \citep[OGLE, ][]{ogleiv} data to search for $\delta$~Scuti stars in the Galactic bulge region, which resulted in the discovery of more than 10\,000 pulsators \citep{pietrukowicz2020}. These stars were then a subject of the study to search for multi-mode radially pulsating HADS \citep{netzel_hads} and resulted in a sample of 2657 candidates for stars pulsating in two radial modes simultaneously, 414 candidates for stars pulsating in three radial modes and 14 candidates for stars pulsating in four radial modes. Petersen diagram of all these stars is presented in fig.~7 of \cite{netzel_hads}. In this study we aim to perform asteroseismic modeling of candidates for pulsations in at least three radial modes in order to obtain basic stellar parameters of these stars, similarly to \cite{pietrukowicz2013}. Our input data consists only of information about the periodicities. Physical parameters, including luminosities and effective temperatures, are determined solely by matching observed periods to theoretical periods from the grid of models.

\section{Data and analysis}

For the asteroseismic modeling we used multimode HADS identified by \cite{netzel_hads} as pulsating in at least three radial modes, i.e. a sample of 428 stars. In particular, there are 145 stars pulsating in fundamental mode, first overtone and second overtone (F+1O+2O); 221 stars pulsating in fundamental mode, first and third overtones (F+1O+3O) and 48 stars pulsating in first, second and third overtones (1O+2O+3O). Fourteen stars are candidates for pulsations in four modes: fundamental mode, first, second and third overtones (F+1O+2O+3O). Due to the numerous sample we did not study these stars individually, but compared their periods with calculated theoretical periods from a grid of models. First, we calculated the grid of evolutionary models using version 11701 of MESA \citep{mesa1,mesa2,mesa3,mesa4,mesa5}. Based on these models, we calculated periods of radial pulsations using the code by \cite{dziembowski1977}.  Then, we selected the models that match the observed periods of $\delta$~Scuti stars best. 

A grid of evolutionary models was computed with mass, $M$, metallicity, [Fe/H], and overshooting from the hydrogen-burning convective core, $f_{\rm ov}$, considered as variable parameters. The range for masses is $0.8-2.5$\,M$_\odot$ with a step of 0.05\,M$_\odot$. Although, as mentioned in Sec.~1, the masses of $\delta$~Scuti stars are typically in the range $1.4-2.5$\,M$_\odot$, we included smaller masses in order to cover the mass range corresponding to SX~Phe stars.
Metallicities, [Fe/H], are in the $-2.0$ to $+0.5$ range, with a step of 0.125\,dex. Mass fractions of hydrogen helium and metals necessary for evolutionary calculations were computed from the [Fe/H] values assuming scaled solar abundances as determined by \cite{asplund2009}. We used OPAL opacities \citep{iglesias1996} supplemented with opacity tables for low temperatures \citep{ferguson2005}. We used photospheric tables for the atmosphere description \citep{hauschildt1999a,hauschildt1999b,castelli.kurucz2003}. We used the exponential scheme for the description of convective overshooting at the border of hydrogen burning convective core \citep{herwig2000}. In the traditional approach, the extent of overshooting, $d_{\rm ov}$, is described as $d_{\rm ov} = \alpha_{\rm ov} H_p$, where $H_p$ is a pressure scale height and $\alpha_{\rm ov}$ is the overshooting parameter. The overshooting parameter $f_{\rm ov}$, used in exponential prescription, corresponds to $\alpha_{\rm ov}$ as $f\approx 0.1 \alpha_{\rm ov}$ \citep{moravveji2015}. In the considered model grid, overshooting parameter, $f_{\rm ov}$, was equal to 0.0 (no overshooting), 0.01, 0.02 or 0.03. Mixing length parameter, $\alpha_{\rm MLT}$, was set to 1.76 based on the solar calibration performed with the calibration algorithm within MESA. The extent of the convection region was determined using the convective premixing algorithm \citep{mesa5}. Effects of rotation and mass loss were neglected. 
Evolutionary tracks were calculated from the zero-age main sequence to the point where star has effective temperature lower than 5500 K, which corresponds to evolution well past the $\delta$~Scuti instability strip. We did not include pre-main-sequence evolution, because that would significantly increase the time of calculations and number of stars expected to be at this stage of evolution is small. The time step between the models on each evolutionary track was not constant. It depends on evolutionary stage and internal structure of the model and is set in MESA through several parameters controlling the time resolution and relative structure variations between the models (see inlist in the Appendix). As a result, different number of models was calculated for different evolutionary tracks. In total, the computed grid of evolutionary tracks consists of 5\,227\,457 models. An example of MESA inlist file is provided in Appendix~\ref{inlist}.

Pulsation modes were determined at each step of the evolutionary tracks using the Warsaw Pulsation Code \citep{dziembowski1977}. Full evolutionary structure was used as an input to pulsation calculations. The pulsation code provided us with periods of radial modes and the instability parameter. The frozen-in assumption for convection is adopted in the code. Hence, the pulsation-convection coupling is neglected and the red edge of the instability strip cannot be determined with this code. The position of the blue edge can be approximately determined. We note however, that pulsation-convection coupling has a small effect on the blue edge as well. We determined the position of the blue edge of the instability strip and used it to constrain the number of possible models in the grid. Since we did not have information about the position of the red edge, after performing the fitting process, we compared positions of fitted models to positions of red edges determined in other studies \citep[e.g.][]{xiong2016}. We did not obtain many solutions that would have significantly lower effective temperatures than corresponding to the the red edge. 

In this analysis we used linear periods of pulsations. The effect of nonlinear shift in frequencies on period ratios is relatively small. Nonlinear period ratios are usually smaller than linear period ratios by a few tenths of a per cent as shown by \cite{kollath.buchler2001} for classical Cepheids and RR Lyrae stars.

For each star, models with periods and period ratios matching the observed ones best, were selected. To accomplish that, we defined the $D$ parameter, which describes the difference between the longest observed ($P_{\rm longest}^{\rm o}$) and calculated (model) period ($P_{\rm longest}^{\rm m}$), and differences between all period ratios possible for a given type of multimode radial pulsations:
\begin{equation}
D^2=\left( \frac{P_{\rm longest}^{\rm m} - P_{\rm longest}^{\rm o}}{P_{\rm longest}^{\rm o}} \right)^2 + \sum_{i,j} \left( \frac{R_{i,j}^{\rm m} - R_{i,j}^{\rm o}}{R_{i,j}^{\rm o}} \right)^2,
\label{Eq.d}
\end{equation}
where $R_{i,j}=P_i/P_j$ are period ratios ($P_i<P_j$). For each star we selected one model from the grid that has the smallest value of the $D$ parameter, denoted as $D_{\rm min}$.
  
We note that we tested other expressions for the $D$ parameter, including the expression used by \cite{pietrukowicz2013}, as well as $\chi^2$:
 \begin{equation}
\chi^2=\sum_i \frac{1}{\sigma_i^2}\left( P_i^{\rm m} - P_i^{\rm o} \right)^2,
\label{Eq.chi}
\end{equation}
where $\sigma_i$ is the error of observed period, $P_i^{\rm o}$. For each of the tested expressions, the parameters of the fitted models are similar and the results and conclusions for the whole sample of $\delta$~Sct stars do not change (see discussion in Sec.~\ref{Sec.discussion}.). The observational errors do not play a significant role for the analyzed stars, since they are small, of the order of $10^{-7}$, for all stars in the sample. We note that due to simplifying assumptions we make in our evolutionary and pulsation modelling, these errors are smaller than accuracy of calculated pulsation periods. We decided to choose the $D$ parameter from Eq.~\ref{Eq.d} for the fitting. 
  
\begin{figure*}
\includegraphics[width=\textwidth]{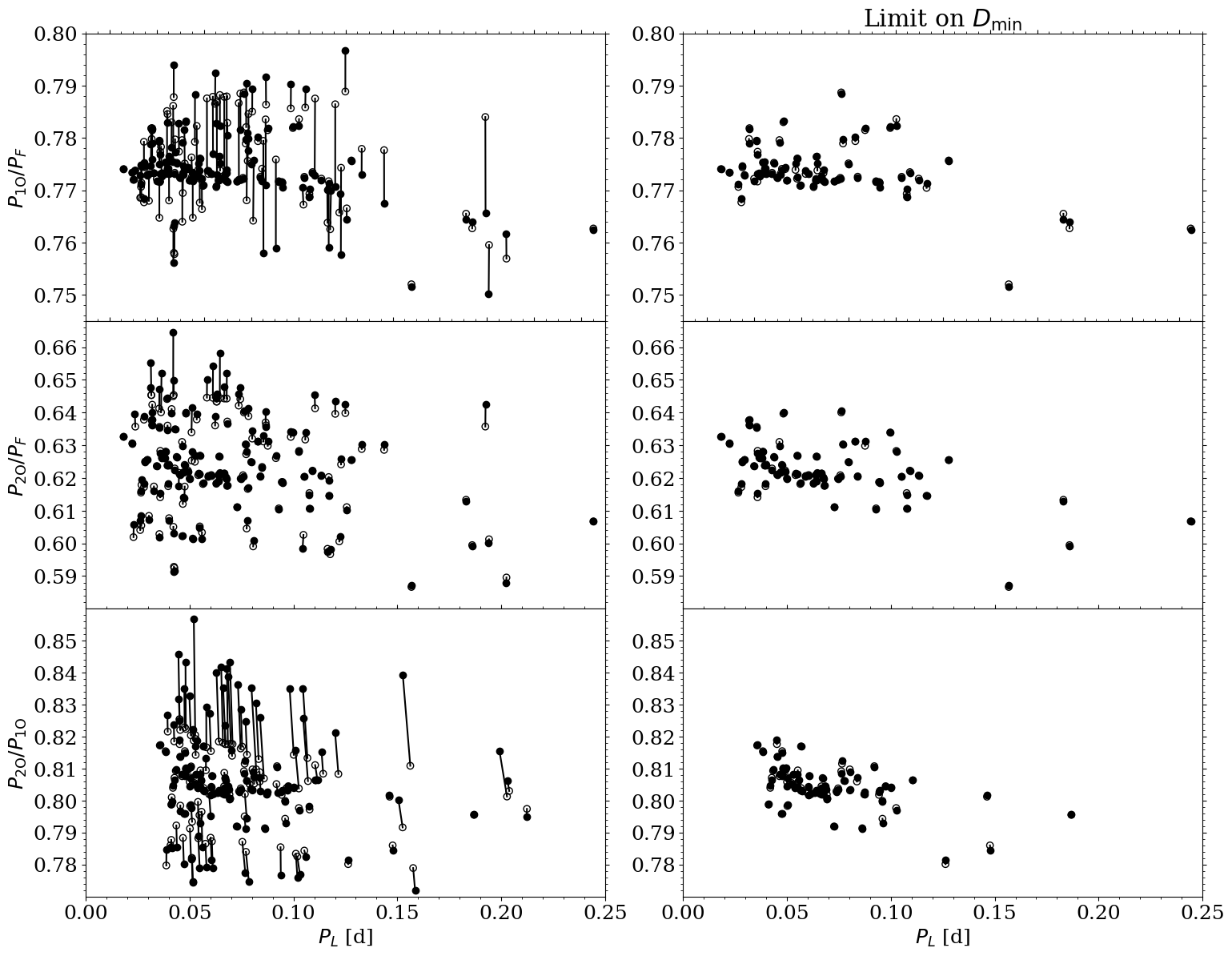}
\caption{Petersen diagrams for the F+1O+2O stars. For earch star, the observed and theoretical values from the model with $D_{\rm min}$ are plotted with filled and open symbols and connected with a line. The left panels show all stars without any limit on the $D_{\rm min}$, whereas the right panels show stars with the parameter $D_{\rm min}\leq 0.003$ (see text).}
\label{Fig.d_f1o2o}
\end{figure*}

\begin{figure}
\includegraphics[width=0.5\textwidth]{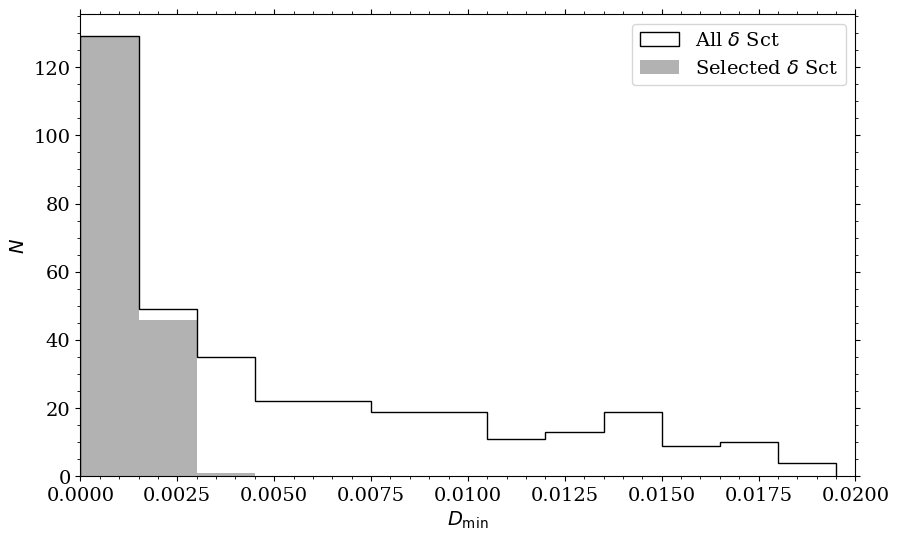}
\caption{Distribution of the $D_{\rm min}$ parameter for all analyzed $\delta$ Scuti stars (black line) and for stars selected based on the Petersen diagrams (gray area).}
\label{Fig.hist_d}
\end{figure}

 \begin{figure*}
\includegraphics[width=\textwidth]{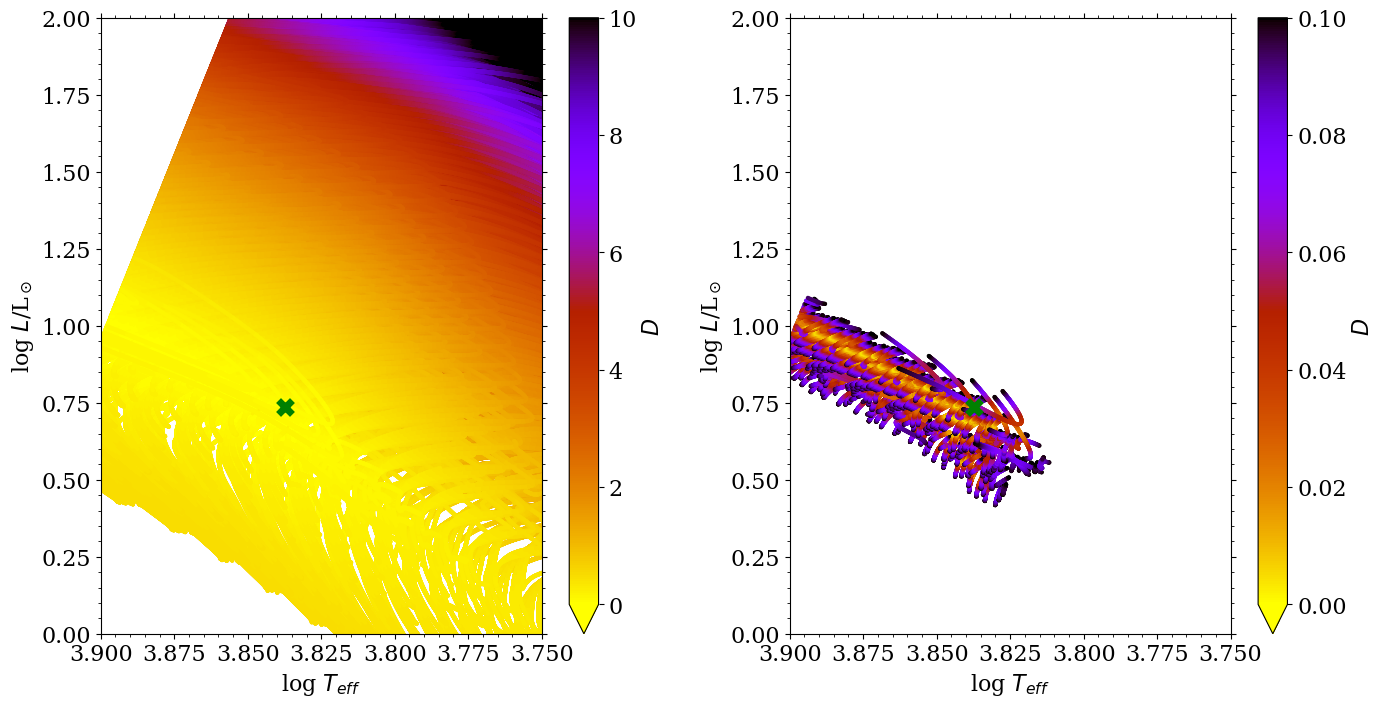}
\caption{The Hertzsprung-Russel diagram with the grid of models. Fit for OGLE-BLG-DSCT-09055 is plotted with a green cross. Values of the $D$ parameter are color coded as indicated in the key. All models are presented in the left panel. Models with $D\leq 0.1$ are plotted in the right panel.}
\label{Fig.d_min}
\end{figure*}

 Even though the calculated grid of models is very dense, the models with the smallest value of the $D$ parameter, $D_{\rm min}$, are not considered good enough for some of the stars. This problem is presented in Fig.~\ref{Fig.d_f1o2o}. We used a group of stars pulsating in F+1O+2O as an example. For each star we plotted two points corresponding to observed (filled symbols) and calculated (open symbols) values of periods and period ratios (that minimize the value of $D$). Both points are connected with a line for each star. As visible in the left panels of Fig.~\ref{Fig.d_f1o2o}, for some stars the differences are significant and the models cannot reproduce the observed values of periods and period ratios. Hence, we decided to introduce a limit on the $D_{\rm min}$ parameter, below which the models can be considered well fitted. We tested different values of the limit on the $D_{\rm min}$ parameter and chose the one that provides well fitted models but does not reduce the number of fitted stars too much so the sample of stars with determined physical parameters is still statistically significant. For the F+1O+2O and F+1O+3O stars we chose this limit to be $D_{\rm min}=0.003$. For the 1O+2O+3O stars we chose this limit to be $D_{\rm min}=0.002$ and for the F+1O+2O+3O stars -- $D_{\rm min}=0.005$. The Petersen diagrams after applying the limiting value of the $D_{\rm min}$ parameter is presented in the right panels of Fig.~\ref{Fig.d_f1o2o}. After excluding stars for which the $D_{\rm min}$ parameter is too big, we reduced the sample of stars with fitted models to 176. In Fig.~\ref{Fig.hist_d} we showed the distribution of the $D_{\rm min}$ parameter for each star from the sample and for stars selected using Petersen diagrams as described above. 
 
 The possible reasons for not finding well fitted models for some of the studied stars are discussed further in Sec.~\ref{Sec.discussion}. We also note, that even though for each star we selected a single model which minimized the $D$ parameter, there were more models with the $D$ parameters of similar order of magnitude. This is illustrated for the F+1O+2O star, OGLE-BLG-DSCT-09055 in Fig.~\ref{Fig.d_min}. We plotted the distribution of the $D$ parameters for models from the grid on the Hertzsprung-Russel diagram. The best matching model for OGLE-BLG-DSCT-09055 is marked with a green cross. The lowest values of the $D$ parameter do not group in a specific region forming a sharp minimum. They rather form a wide region with fits with the $D$ parameters of similar order of magnitude. This suggests that the adopted method is not appropriate for precise asteroseismic modeling of individual stars, but it is still useful for statistical studies of numerous samples of stars (see the comparison of the studied sample with other studies of HADS in Sec.~\ref{Sec.comparison}).

  \begin{table*}
  \centering
     \caption{Sample of a table with observed and calculated periods and period ratios for HADS. For each stars first row gives calculated values and second row gives observed values. Last column provide the parameter $D_{\rm min}$. Full table is available online as supplementary material.}
  \begin{tabular}{lccccccccc}

  \hline
ID		& $P_{\rm L}$ & 	$P_{\rm 1O}$/$P_{\rm F}$ 	&	$P_{\rm 2O}$/$P_{\rm F}$ 	&	$P_{\rm 3O}$/$P_{\rm F}$	&	$P_{\rm 2O}$/$P_{\rm 1O}$ 	&	$P_{\rm 3O}$/$P_{\rm 1O}$	& 	$P_{\rm 3O}$/$P_{\rm 2O}$ & $D_{\rm min}$ \\
\hline
OGLE-BLG-DSCT-00018	&	0.0778	&	0.7782	&	0.6305	& &	0.8102	& & &	0.001548	\\
&	0.0778	&	0.7791	&	0.6307	& &	0.8095	& & & \\
OGLE-BLG-DSCT-00075	&	0.1071	&	0.7815	&	0.6298	& &	0.8059	& & &	0.002742	\\
&	0.1071	&	0.7818	&	0.6311	& &	0.8073	& & & \\
OGLE-BLG-DSCT-00358	&	0.0606	&	0.7720	& &	0.5117	& &	0.6629	& &	0.002709	\\
&	0.0606	&	0.7732	& &	0.5114	& &	0.6615	& & \\
OGLE-BLG-DSCT-00394	&	0.1425	&	0.7757	&	0.6254	& &	0.8063	& & &	0.000134	\\
&	0.1425	&	0.7757	&	0.6255	& &	0.8064	& & & \\
OGLE-BLG-DSCT-00434	&	0.2062	& & & &	0.7942	&	0.6582	&	0.8287	&	0.000281	\\
&	0.2061	& & & &	0.7941	&	0.6581	&	0.8288	& \\
\vdots &	\vdots 	&	\vdots 	&	\vdots 	& \vdots 	&	\vdots 	& \vdots 	& \vdots 	&	\vdots 	\\			
\hline
  \end{tabular}

 \label{tab.f1o2o_d}
 \end{table*}
 

\section{Results}

For 176 stars a satisfactory solution has been found. The resultant sample consists of 68 stars pulsating in F+1O+2O, 81 stars pulsating in F+1O+3O, 25 stars pulsating in 1O+2O+3O and 2 stars pulsating in F+1O+2O+3O. We compare observed period (the longest one) and period ratios with those corresponding to the best fitted models in Table~\ref{tab.f1o2o_d}. The last column of the table provides $D_{\rm min}$.

Physical parameters obtained from the best fitted model for each star are collected in Table~\ref{tab.4mody_sample}. Consecutive columns provide star's ID, type of multimode radial pulsations, mass, metallicity, overshooting parameter, luminosity, effective temperature and age. Remarks regarding evolutionary status are in the last column. Evolutionary status was determined based on central hydrogen abundance in the chosen model. The model is considered to be on the main sequence (MS) if the central hydrogen abundance is larger that $10^{-5}$ and to be post main sequence (PostMS) otherwise. The number of stars on the main sequence is 99, which constitutes 56 per cent of the sample. The remaining 77 stars (44 per cent) are at the evolution stage after the main sequence. 
Considering different types of pulsations, 56 per cent of F+1O+2O stars, 62 per cent of F+1O+3O stars and 36 per cent of 1O+2O+3O stars were classified as main sequence stars. Both stars pulsating in four modes are classified as main sequence stars. In general, the star spends significantly more time during the evolution on the main sequence than at the post-main sequence stage. However, in the considered mass range ($0.8-2.5$\,M$_{\odot}$), not all of these stars have their main-sequence evolution inside of the instability strip. As a results, almost twice as many stars cross the instability strip only during the post-main sequence evolution. This high ratio of post-main sequence stars was also observed in the results of modeling $\delta$~Scuti stars from the Galactic disk \citep{pietrukowicz2013}, who classified 4 out of 7 stars as PostMS.

  \begin{table*}
  \centering
     \caption{Sample of a table with the best models selected for HADS stars pulsating in at least three radial modes simultaneously. Consecutive columns provide star's ID, mass, metallicity, overshooting parameter, age, luminosity and effective temperature. Remarks regarding evolutionary status are in the last column. Full table is available online as supplementary material.}
  \begin{tabular}{ccccccccc}

  \hline
ID & type & 	$M$ [M$_\odot$]	&	[Fe/H]	&	$f_{\rm ov}$	&	log $L$ [L$_\odot$]&	log $T_{\rm eff}$ [K] 	& Age [Gyr] &	Remarks	\\
\hline
OGLE-BLG-DSCT-00018	&	F+1O+2O	&	1.20	&	--0.875	&	0.00	&	3.87404	&	1.01981	&	3.14	&	PostMS	\\
OGLE-BLG-DSCT-00075	&	F+1O+2O	&	1.05	&	--1.375	&	0.00	&	3.83517	&	0.99743	&	4.72	&	PostMS	\\
OGLE-BLG-DSCT-00358	&	F+1O+3O	&	1.10	&	--0.625	&	0.00	&	3.84097	&	0.69262	&	4.14	&	MS	\\
OGLE-BLG-DSCT-00394	&	F+1O+2O	&	1.55	&	--0.750	&	0.00	&	3.86804	&	1.41264	&	1.50	&	PostMS	\\
OGLE-BLG-DSCT-00434	&	1O+2O+3O	&	2.50	&	0.250	&	0.03	&	3.85956	&	1.86218	&	0.75	&	MS	\\
OGLE-BLG-DSCT-00648	&	F+1O+2O	&	1.65	&	0.125	&	0.00	&	3.86256	&	0.93507	&	0.97	&	MS	\\
OGLE-BLG-DSCT-00871	&	F+1O+3O	&	1.10	&	--1.125	&	0.00	&	3.87732	&	0.93690	&	4.01	&	PostMS	\\
OGLE-BLG-DSCT-00899	&	F+1O+3O	&	1.35	&	--0.500	&	0.00	&	3.84224	&	1.08552	&	2.42	&	PostMS	\\
\vdots	&	\vdots	&	\vdots	& \vdots	&	\vdots	&	\vdots	& \vdots &	\vdots	&	\vdots	\\
\hline
  \end{tabular}

 \label{tab.4mody_sample}
 \end{table*}

\begin{figure*}
\includegraphics[width=\textwidth]{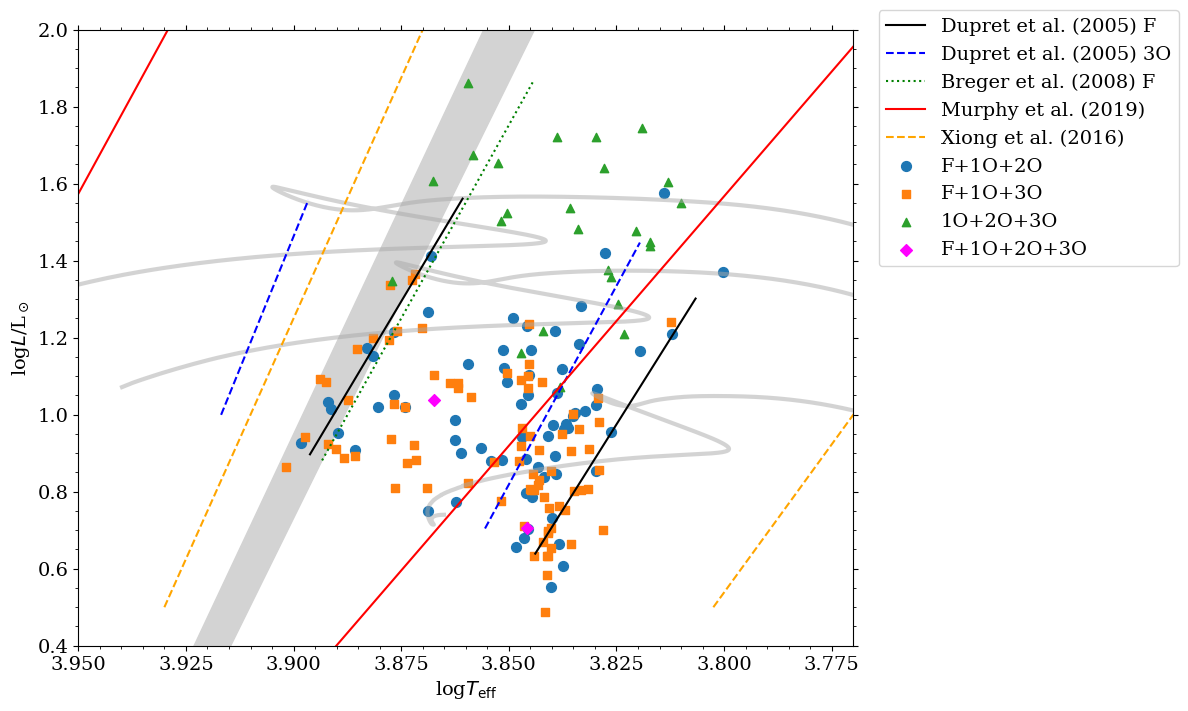}
\caption{HR diagram showing HADS with determined physical parameters. Different markers correspond to different types of multimode pulsations: blue circles - F+1O+2O, orange squares - F+1O+3O, green triangles - 1O+2O+3O and red diamonds - F+1O+2O+3O. With different lines we included edges of the instability strip from the literature, as indicated in the key and described in the text. Grey area corresponds to the range of blue edges for different metallicities from this study.}
\label{Fig.fitted}
\end{figure*}

On the Hertzsprung-Russel diagram in Fig.~\ref{Fig.fitted} we plotted stars with physical parameters determined together with theoretical and observational instability strips. Different markers correspond to different types of multimode radial pulsations as described in the key and the caption. The grey area shows the range of blue edges determined in this study from pulsation models for different metallicities, hence all stars are located towards the lower temperature than the left edge of the area. Green dotted line corresponds to the theoretical blue edge for the fundamental mode determined by \cite{breger2008}, who also could not determine the position of the red edge. Red solid lines correspond to the blue and red edges determined based on observations by \cite{murphy2019} who used $\delta$~Scuti stars observed by {\it Kepler}. As visible from Fig.~\ref{Fig.fitted}, the full sample of $\delta$~Scuti stars observed by {\it Kepler} covers a wider range of effective temperatures than radially pulsating HADS from this study. \cite{dupret2005}, using time-dependent prescription for convection, calculated positions of the blue and red edges, which are plotted with black solid lines for the fundamental mode and with blue dashed line for the third overtone. Orange dotted lines correspond to theoretically predicted positions of the blue and red edges for low order radial modes by \cite{xiong2016}, who used a non-local and time-dependent treatment of convection.

Stars pulsating in F+1O+2O, F+1O+3O and F+1O+2O+3O cover similar ranges in effective temperature and luminosity. With an exception of one F+1O+2O star they do not exceed 1.5 in $\log L/{\rm L}_{\odot}$. There is a grouping of stars at the low-luminosity and low-temperature part of the Herzsprung-Russel diagram, however still within the boundaries of the instability strip predicted by theoretical studies \citep{xiong2016}. These are typically low mass stars. Stars pulsating in 1O+2O+3O have, on average, significantly higher luminosity than the rest of the sample. They also do not group at any particular edge of the instability strip.

\begin{figure}
\includegraphics[width=0.5\textwidth]{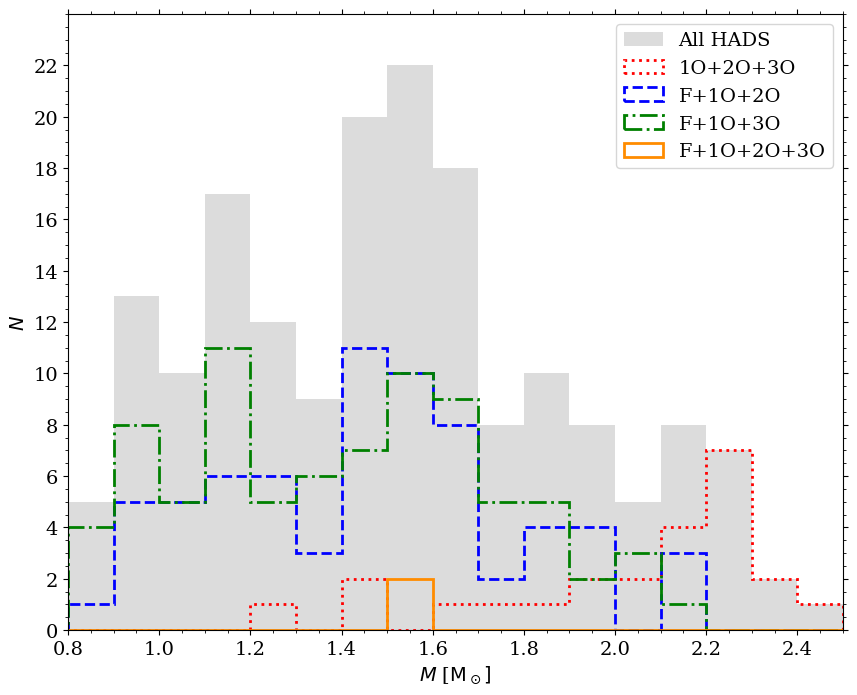}
\caption{Distribution of mass. Different lines and colors mark different types of multimode pulsations as indicated in the key.}
\label{Fig.m_histogram}
\end{figure}

In Fig.~\ref{Fig.m_histogram} we plotted the distribution of mass with distinction to different types of multimode pulsations. Two stars that pulsate in four modes both have masses 1.55\,M$_{\odot}$. Triple-mode stars cover whole range of masses from the grid. Majority of them however are below 2\,M$_{\odot}$. The most numerous among low mass stars, with masses below 1\,M$_{\odot}$, are that pulsating in F+1O+3O. Stars pulsating in F+1O+2O, as well as in F+1O+3O, cover a wide range of masses from 0.8\,M$_{\odot}$ to 2.2\,M$_{\odot}$. Stars pulsating in 1O+2O+3O are less numerous in the low mass range. The lowest mass in this group is 1.25\,M$_{\odot}$. They are the most numerous group of massive stars, which is consistent with their location in the HR diagram (Fig.~\ref{Fig.fitted}). Majority of them have masses above 2\,M$_{\odot}$. Only a few triple-mode stars with the fundamental mode excited, have masses above 2\,M$_{\odot}$. The maximum of the distribution for all stars, regardless of the type of pulsation, corresponds to masses $1.4-1.7$\,M$_{\odot}$.

\begin{figure}
\includegraphics[width=0.5\textwidth]{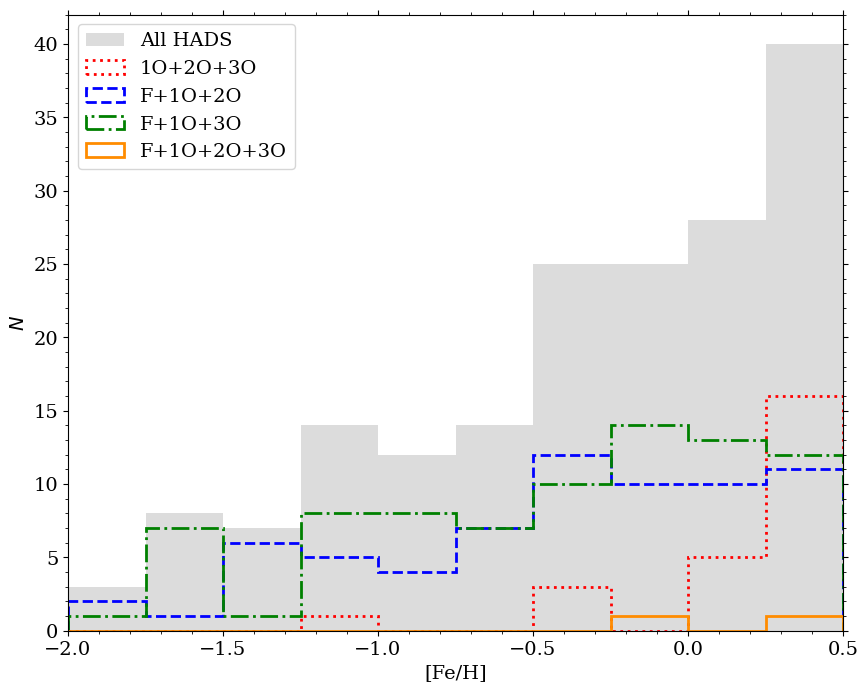}
\caption{Distribution of metallicity. Different lines and colors mark different types of multimode pulsations as indicated in the key.}
\label{Fig.feh_histogram}
\end{figure}

In Fig.~\ref{Fig.feh_histogram} we have plotted distribution of metallicity with distinction for different pulsation types. Again, studied stars cover whole range of metallicity possible in the calculated grid. The number of stars increases with increasing metallicity. The most common metallicity among all types is close to solar, between ${\rm [Fe/H]}=-0.5$ and ${\rm [Fe/H]}=0.5$. Two stars pulsating in four modes have metallicities of ${\rm [Fe/H]}=-0.25$ and ${\rm [Fe/H]}=+0.25$. Stars pulsating in 1O+2O+3O have typically higher metallicities than other considered types. They constitute the majority of stars in the highest metallicity bin of the distribution. Stars with low metallicities, below $-0.75$, are less frequent. These low-metallicity stars are either F+1O+2O, or F+1O+3O. 

\begin{figure}
\includegraphics[width=0.5\textwidth]{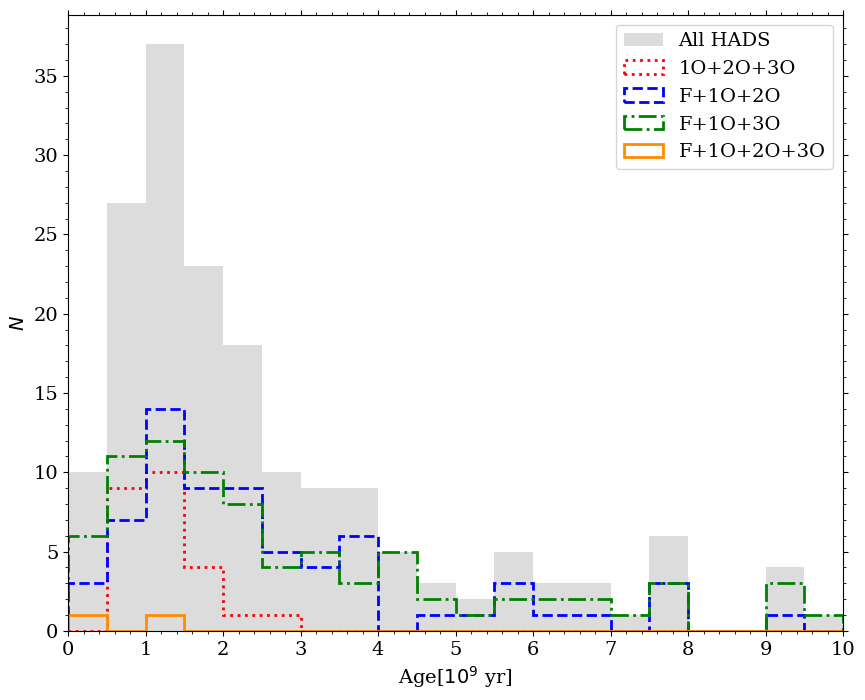}
\caption{Distribution of age. Different lines and colors mark different types of multimode pulsations as indicated in the key.}
\label{Fig.age_histogram}
\end{figure}

In Fig.~\ref{Fig.age_histogram} we have plotted distribution of ages in the sample. Median value of age for the whole sample is 1.7 Gyr. A tail of older stars, with ages up to 10 Gyr, is also visible. Among stars with high ages, above 3 Gyr, there are only stars pulsating in F+1O+2O and F+1O+3O. Median value of age for stars pulsating in 1O+2O+3O is around 1.1 Gyr and the distribution is not as wide as for the rest of the stars.

\begin{figure}
\includegraphics[width=0.5\textwidth]{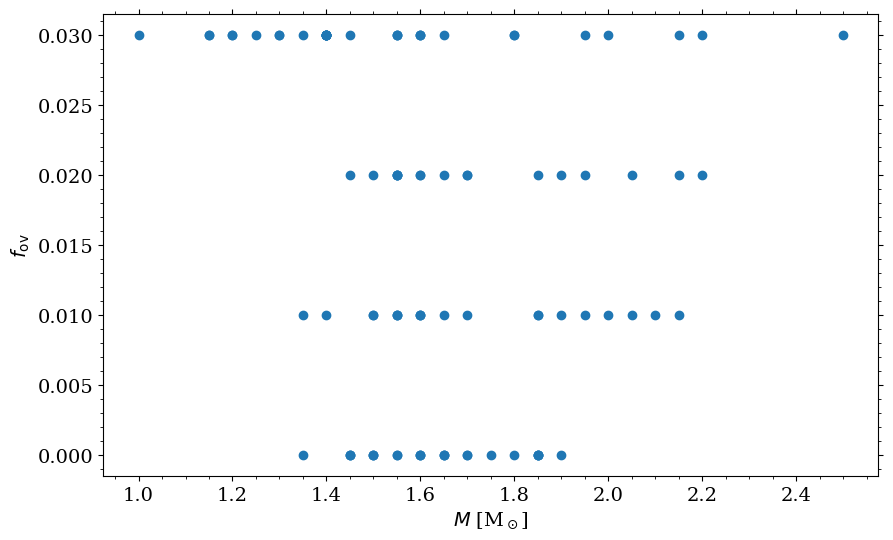}
\caption{Relation between the overshooting parameter and mass for fitted models with a convective cores.}
\label{Fig.fov}
\end{figure}

\begin{figure}
\includegraphics[width=0.5\textwidth]{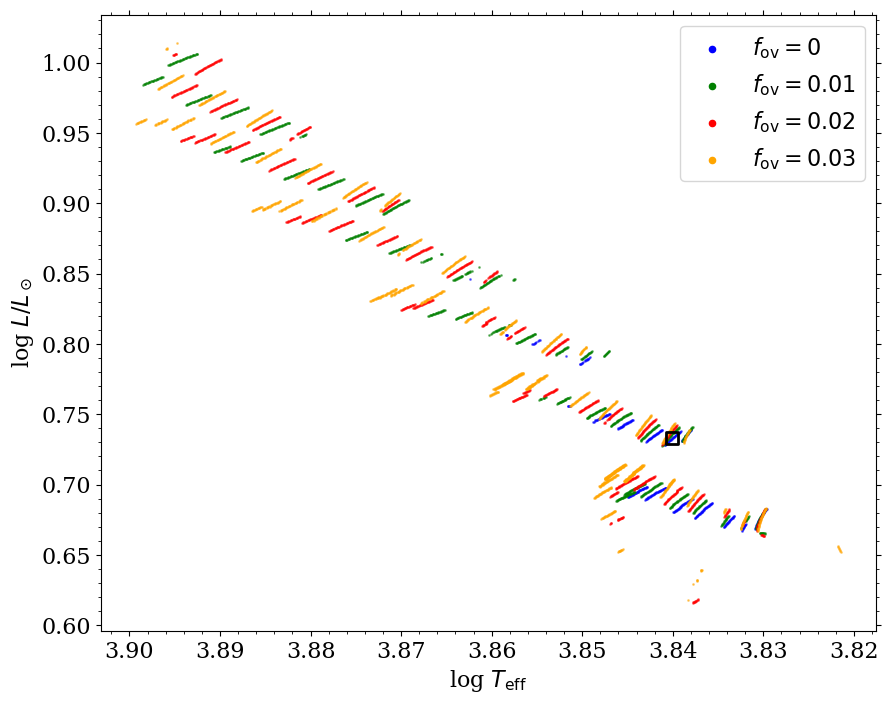}
\caption{Hertzsprung-Russel diagam with models having convective core and $D \leqslant 0.02$ for OGLE-BLG-DSCT-09055. Models with different values of the convective overshooting parameter are plotted with different colors as indicated in the key. The model fitted to OGLE-BLG-DSCT-09055 is plotted with black square.}
\label{Fig.fov_09055}
\end{figure}

In Fig.~\ref{Fig.fov} we plotted relation between the overshooting parameter and the mass for the fitted models with a convective core. The relation of increasing overshooting parameter with increasing mass is discussed in literature \citep[see e.g.][and references therein]{constantino.baraffe2018}. However, no clear relation between the mass and the convective overshooting parameter is visible for the models analyzed here. In particular, there are models with masses smaller than 1.4 M$_\odot$ with the highest value of the overshooting parameter ($f_{\rm ov}=0.03$). Similar effect was observed by \cite{deheuvels2016} who noticed that for high values of the overshooting parameters even low-mass stars develop the convective core. We also note, that fitting of the overshooting parameter is not reliable. In the Hertzsprung-Russel diagram in Fig.~\ref{Fig.fov_09055}, we plotted models with convective core present and $D \leqslant 0.02$ for OGLE-BLG-DSCT-09055 (that is also plotted in Fig.~\ref{Fig.d_min}). With different colors we plotted different values of the overshooting parameter. The best model fitted to the OGLE-BLG-DSCT-09055 is marked with a black square. As visible in Fig.~\ref{Fig.fov_09055}, models with different overshooting parameters are close to the best model fitted to OGLE-BLG-DSCT-09055, which means that the parameter $D$ has similar values for all of them (compare with Fig.~\ref{Fig.d_min}). The method of the fitting based on periods and period ratios alone, do not allow for unambiguous determination of the overshooting parameter.

\begin{figure}
\includegraphics[width=0.5\textwidth]{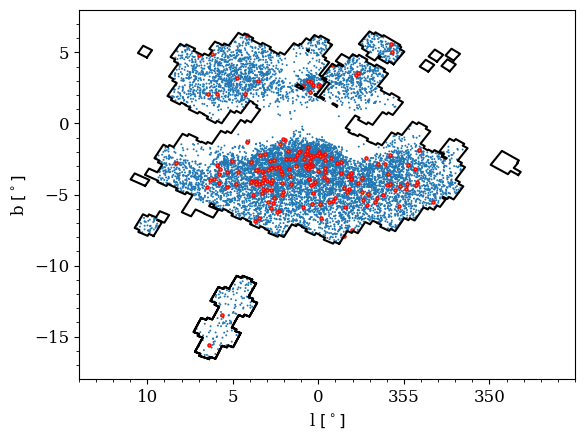}
\caption{Location of HADS from this study together with all $\delta$~Scuti stars identified in the OGLE Galactic bulge fields \citep{pietrukowicz2020}.}
\label{Fig.locations}
\end{figure}

In Fig.~\ref{Fig.locations} we plotted locations of HADS from this study together with all $\delta$~Scuti stars identified in the OGLE Galactic bulge fields \citep{pietrukowicz2020}. Distribution of multimode HADS follows the overall distribution of all $\delta$~Scuti stars.

\begin{figure*}
\includegraphics[width=\textwidth]{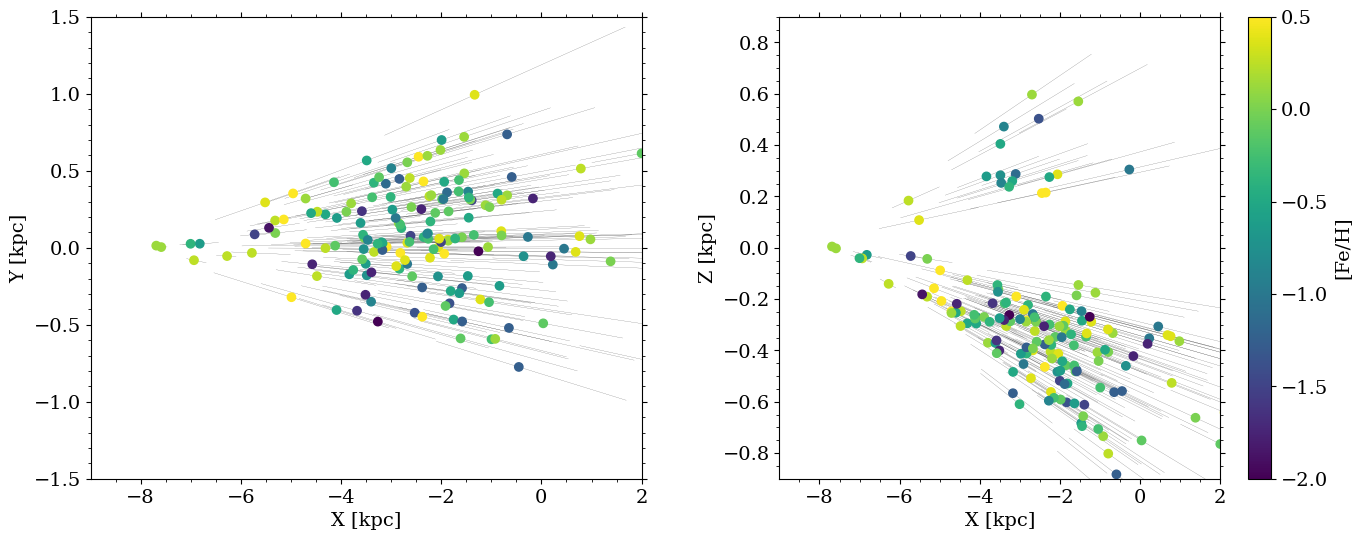}
\caption{Positions of HADS from the analyzed sample in the galactocentric coordinates. The top view is presented in the left panel, the side view is presented in the right panel. Color corresponds to the metallicity as indicated in the key. Gray lines correspond to possible positions based on the 16th and 84th percentile of the distance determination in the Gaia EDR2 data.}
\label{Fig.mapka_gal}
\end{figure*}

We used Gaia EDR3 distances and positions \citep{edr2021,edr3_distances} available for HADS from our sample to determine their positions in galactocentric coordinates. Either geometric or photogeometric distances are available for 163 stars. Positions of HADS in galactocentric coordinates are presented in Fig.~\ref{Fig.mapka_gal}. Colors correspond to metallicities determined in this study. We do not see any clear relation of metallicity or other physical parameters with location of stars. Note, that stars studied in this sample cover relatively small part of the Galactic disk. Possible relations between metallicity and position in the Galactic disk might not be detectable with the considered sample \citep[see e.g.][]{loktin.popova2020}. Additionally, the range of position determination for the studied stars is significant.


\section{Discussion}\label{Sec.discussion}

\subsection{Fitting the models} 

\begin{figure}
\includegraphics[width=0.5\textwidth]{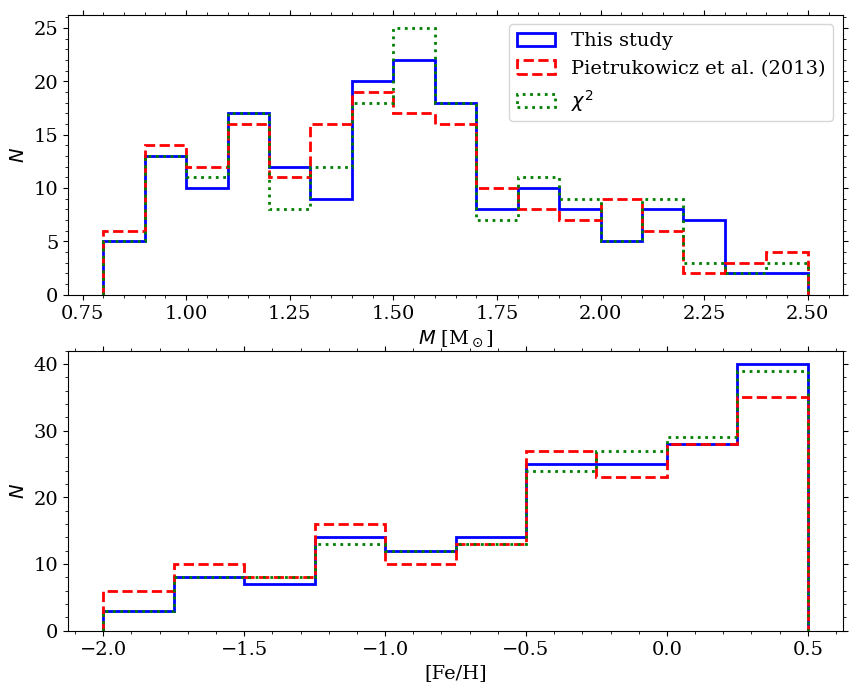}
\caption{Distributions of mass (top panel) and metallicity (bottom panel) resulting from minimization of three different parameters: $D$ adopted in this study (Eq.~\ref{Eq.d}; blue solid line), a parameter used by \protect\cite{pietrukowicz2013} (red dashed line) and $\chi^2$ (Eq.~\ref{Eq.chi}; green dotted line).}
\label{Fig.comp_d_feh_m}
\end{figure}

We selected well fitted models by minimizing the $D$ parameter (Eq.~\ref{Eq.d}), which is a modified version of the parameter used for model fitting by \cite{pietrukowicz2013}, and puts more emphasis on matching the period ratios. However, to make sure that results are essentially the same when using different expressions for the model fitting, for stars with well fitted models we have repeated the fitting procedure by minimizing the values of the parameter used by \cite{pietrukowicz2013} and of $\chi^2$ (Eq.~\ref{Eq.chi}). In Fig.~\ref{Fig.comp_d_feh_m} we plotted the resulting distributions of the mass and metallicity. The differences for individual stars are small, if any, and the distributions of physical parameters are qualitatively the same for all three fitting methods.

Good fits to the observed period ratios were possible for 176 multimode HADS stars and hence their physical parameters were determined. This is the most numerous and homogeneous group of HADS studied with asteroseismic methods so far. The input sample consists of over 400 stars, however we did not find good fits for the majority of them (see left panels of Fig.~\ref{Fig.d_f1o2o}). This might be due to the fact that multi-periodic radial HADS selected by \cite{netzel_hads} were classified based only on period ratios. In the case of some stars the classification might be incorrect. For example, considering triple-mode F+1O+2O candidates, in some stars only one period ratio could be fitted correctly and for the other period ratios we observed a significant difference between observed and fitted values. As visible in Fig.~\ref{Fig.d_f1o2o}, the differences are typically smaller for period ratio $P_{\rm 2O}/P_{\rm F}$ than for the other two period ratios. Some of these stars might be in fact double-mode F+2O $\delta$~Scuti stars. The remaining signal, originally interpreted as the first overtone, might be in fact a non-radial mode. Such stars (this problem may affect all types of modelled multi-mode HADS) would require further careful study and ideally an independent confirmation of radial nature of the observed signals. 

The grid used for fitting was constrained to masses from a range $0.8 - 2.5 $ M$_\odot$ and metallicities from $-2.0$ to $+0.5$\,dex. It is possible that some of the stars for which we did not find a well fitted models have parameters outside the ranges used in the grid. In particular, including masses higher than 2.5\,M$_\odot$ might results in fitting additional stars.

Another factor that may influence differences between observed period ratios and calculated period ratios is rotation that was neglected in the present analysis. Rotation effects of second order modify frequencies of radial modes and in turn modify observed period ratios. \cite{perez1995} calculated theoretical period ratios for radial mode without and with rotation of $v_{\rm rot}=10\,\mu{\rm Hz}$ and $v_{\rm rot}=20\,\mu{\rm Hz}$. The effect on period ratios is the strongest for low radial orders. The differences introduced by rotation are not big enough to influence identification of the modes based on period ratios. However, differences may be significant enough, compared to differences between observed and calculated periods in this analysis. In the case of some stars the lack of well fitted model might be a result of their faster rotation. However, without the information on the rotation rates we cannot check this possibility in detail. HADS have typically lower rotation rates than low-amplitude $\delta$~Scuti stars \citep{breger2000}. Still, they can reach up to $v_{\rm rot}=6.9\,\mu{\rm Hz}$. Impact of rotation, especially for longer periods, which are reached by stars in this sample, can lead to possible differences between models and observations. This aspect is an interesting path for the further study.

\subsection{SX~Phe stars}
SX~Phe stars are typically defined as population II low-metallicity counterparts of $\delta$~Scuti stars \citep[e.g.][]{breger2000}. Their pulsation properties are similar to those of HADS, with typically one or two radial modes excited, however, non-radial and low-amplitude pulsations are also possible \citep[e.g.][]{olech2005}. In globular clusters and old open clusters they are identified as blue stragglers \citep[e.g.][]{olech2005} and in the field they are identified based on their kinematic properties \citep{balona2012}. As discussed by \cite{balona2012}, low metallicity alone is sometimes not enough to distinguish between normal $\delta$~Scuti stars and SX~Phe stars, especially in stellar systems other than Milky Way. Indeed, thorough analysis of spectroscopic and photometric data for candidates for SX~Phe stars in the {\it Kepler} field \cite{nemec2017} showed that the mean metallicity of the sample is near solar, but the kinematic properties still allowed to identify these stars as members of the old halo population.

\cite{pietrukowicz2013} analyzed 7 HADS and classified 2 as SX~Phe based on their lower metallicity ($Z=0.00044$ and $Z=0.00046$), higher ages (5.68\,Gyr and 4.36\,Gyr) and lower masses ($M=0.94\,{\rm M}_\odot$ and $M=1.01\,{\rm M}_\odot$) than for the rest of their sample. \cite{olech2005} analyzed a sample of SX~Phe stars in the globular cluster $\omega$~Centauri and derived masses in the range of $0.9-1.15$\,M$_\odot$ based on comparison with theoretical pulsation models. \cite{antoci2019} analyzed TESS data for the SX~Phoenicis star, the prototype of the class. Based on the two dominant signals in its frequency spectrum, which correspond to the radial fundamental and first overtone modes, \cite{antoci2019} carried out asteroseismic modeling of this star and obtained the mass of $M=1.05\,{\rm M}_\odot$.

\cite{daszynska2020} analyzed the SX Phoenicis itself. First, they confirmed the radial nature of the modes from photometric amplitudes and phases and then performed seismic modeling to obtain physical parameters, i.e. mass, hydrogen and metal abundances, radius and age. Luminosity and effective temperature from the literature were used as an additional constraint. The best models reproducing observed periods have the metal abundance $Z=0.002$ and hydrogen abundance in $0.67-0.7$ range. Masses are in the range $1.05-1.08$\,M$_\odot$, which is consistent with results of \cite{antoci2019}, and ages are in $2.8-3.07$\,Gyr range. All these models are in the post main sequence stage of evolution. Different values of resultant parameters were obtained when different opacity tables were used.

In this study we also tried to select a sample of candidates for SX~Phe stars. In this attempt, we selected stars with higher ages, lower metallicities and lower masses. The limit on metallicity was set to $-1.5$, for age to 3\,Gyr and for mass to 1.1\,M$_\odot$. This resulted in 16 stars classified as SX~Phe. These stars are marked with `SXPhe' in the remarks column of Table~\ref{tab.4mody_sample} in the supplementary material. All of them are post main sequence stars. 
As discussed by \cite{balona2012}, candidates for SX~Phe stars can be selected based on kinematic properties. Field stars that have high proper motions are expected to belong to thick disk or halo population, hence they are in advanced evolution stages. In the case of stars considered in this sample, we look in the direction of the Galactic bulge, but majority of these stars are located in the disk. We performed similar analysis of proper motions to search for stars that are outliers from the mean proper motions. For 160 stars from our sample, including 14 candidates for SX~Phe selected above, we obtained proper motions from the Gaia EDR3 \citep{gaia2016, gaia2021, lindegren2021}. Additionally, we obtained proper motions for 8\,450 $\delta$~Sct stars from \cite{pietrukowicz2020}. We compared proper motions of HADS and candidates for SX~Phe stars to all $\delta$~Scuti stars. We also compared values of tangential velocities for all three samples. We do not see any outliers in the SX~Phe sample {\color{red}} from the HADS sample or from the sample of all $\delta$~Sct stars. We note, that errors of the proper motions are significant for the $\delta$ Scuti sample, which makes the identification of the outliers uncertain.

\subsection{Comparison with other studies}\label{Sec.comparison}

It would be interesting to compare the physical parameters derived from a purely asteroseismic approach with the independent measurements based on the spectroscopic observations. Independent determination of metallicity or effective temperature and luminosity, would also be helpful in detailed modeling of individual objects as it would provide constraints on the model grid. Unfortunately, for stars from the studied sample, spectroscopic observations are not available. We note, that these stars are located in the OGLE inner bulge fields and their average brightness is around 16 mag in {\it V} band, which would make spectroscopic observations challenging.

\cite{murphy2019} analyzed data from {\it Kepler} for over 15\,000 A- and F-type stars and selected over 1\,900 $\delta$~Scuti stars. For these stars they collected a list of physical parameters, i.e. luminosities or effective temperatures. Based on this observed sample, they derived the blue and red edges for $\delta$~Scuti pulsations that we also included in Fig.~\ref{Fig.fitted}. Note, that they did not distinguish between HADS and other $\delta$~Scuti stars.

\cite{pietrukowicz2013} studied stars in the Galactic disk fields from the OGLE~III data and classified 58 stars as $\delta$~Scuti. In 28 stars only one frequency was detected. Out of 30 multiperiodic stars they found, two radial modes were detected in 22 stars and three radial modes in 6 stars. In one star they detected four radial modes simultaneously. Using similar approach to described in this study, they estimated physical parameters of these seven multiperiodic radially pulsating HADS and provided masses, effective temperatures, distance moduli, metal content and ages. Based on these, they classified two stars as SX~Phe.

\cite{mcnamara2000} combined a list of 26 HADS for which they derived physical parameters based on the Str\"{o}mgren photometry. We checked this list using the classification from the SIMBAD database \citep{simbad}. This resulted in a list of 18 HADS and 7 SX~Phe (one star was excluded as classified as RR Lyr in the SIMBAD database). The parameters obtained by \cite{mcnamara2000} are, among others, masses, ages and metallicity.

\begin{figure}
\includegraphics[width=0.5\textwidth]{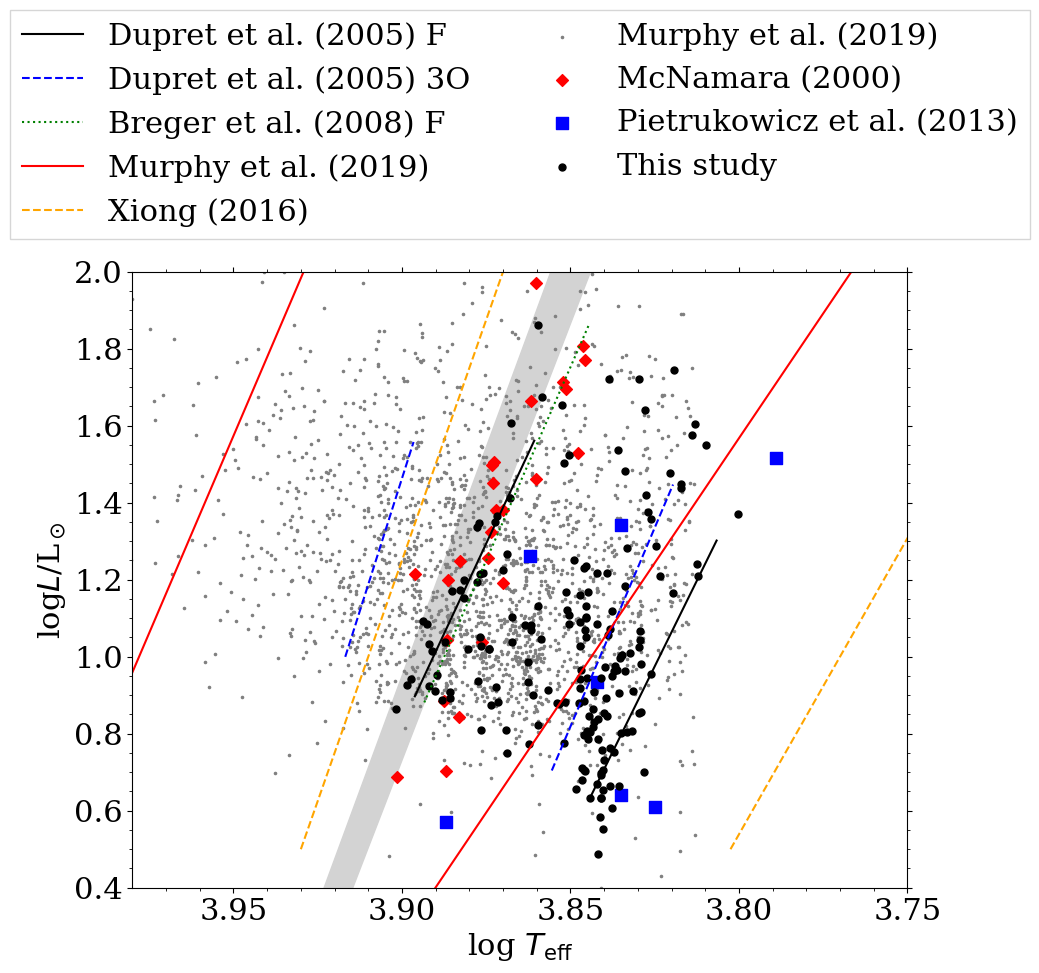}
\caption{Hertzsprung-Russel diagram with HADS from this study plotted with black circles, $\delta$~Scuti stars from \protect\cite{murphy2019} -- grey points, HADS from \protect\cite{pietrukowicz2013}  -- blue squares, and HADS from \protect\cite{mcnamara2000} -- red diamonds. Grey are are positions of the blue edge determined in this study for different metallicities. Other blue and red edges were determined either observationally or theoretically in the literature as indicated in the key.}
\label{Fig.comparison}
\end{figure}

On the Hertzsprung-Russel diagram in Fig.~\ref{Fig.comparison}, we compared the position of modeled HADS studied here to the $\delta$~Scuti sample from the \cite{murphy2019} and both samples of HADS studied by \cite{mcnamara2000} and \cite{pietrukowicz2013}. $\delta$~Scuti stars from \cite{murphy2019} are plotted with grey points, whereas HADS from this work are plotted with black circles. HADS stars studied by \cite{mcnamara2000} are plotted with red diamonds and the sample studied by \cite{pietrukowicz2013} is plotted with blue squares. In Fig.~\ref{Fig.comparison} we provided also different positions of blue and red edges of the instability strip; same as in Fig.~\ref{Fig.fitted}. The whole sample of $\delta$~Scuti stars from {\it Kepler}  covers the widest region on the Hertzsprung-Russel diagram. HADS are placed inside a narrow strip, which width is much smaller than for the rest of $\delta$~Scuti stars and which is located close to the center of the instability strip derived for all $\delta$~Scuti stars.  Especially, HADS from \cite{mcnamara2000} form a narrow strip close the center of the instability strip determined by \cite{murphy2019}. They are also very close to the blue edge determined in this study, as well as by \cite{xiong2016} who determined position of the blue and red edge theoretically. Similarly, \cite{balona2016} used the sample collected by \cite{rodriguez2000} and classified stars with amplitudes larger than 0.3 mag as HADS. They also observed that selected HADS tend to be located in a narrow strip in the middle of all $\delta$~Scuti stars sample in the Hertzsprung-Russel diagram \citep[see the top panel in fig. 1 in][]{balona2016}. Interestingly, \cite{balona2016} did not notice this for $\delta$~Scuti stars based on the data from \cite{poleski2010} \citep[see the bottom panel in fig. 1 in][]{balona2016}. The HADS sample studied by \cite{pietrukowicz2013} cover a wider region and are placed more towards the lower temperatures. Four of them are placed outside the red edge determined for {\it Kepler} $\delta$~Scuti stars, however they are still within the red edge for the fundamental mode as determined by \cite{xiong2016}. HADS in this study are obviously constrained by the blue edge determined here and also extend past the red edge from \cite{murphy2019} but do not reach the red edge from \cite{xiong2016}. In luminosity, stars from all considered samples cover a similar range.

\begin{figure}
\includegraphics[width=0.5\textwidth]{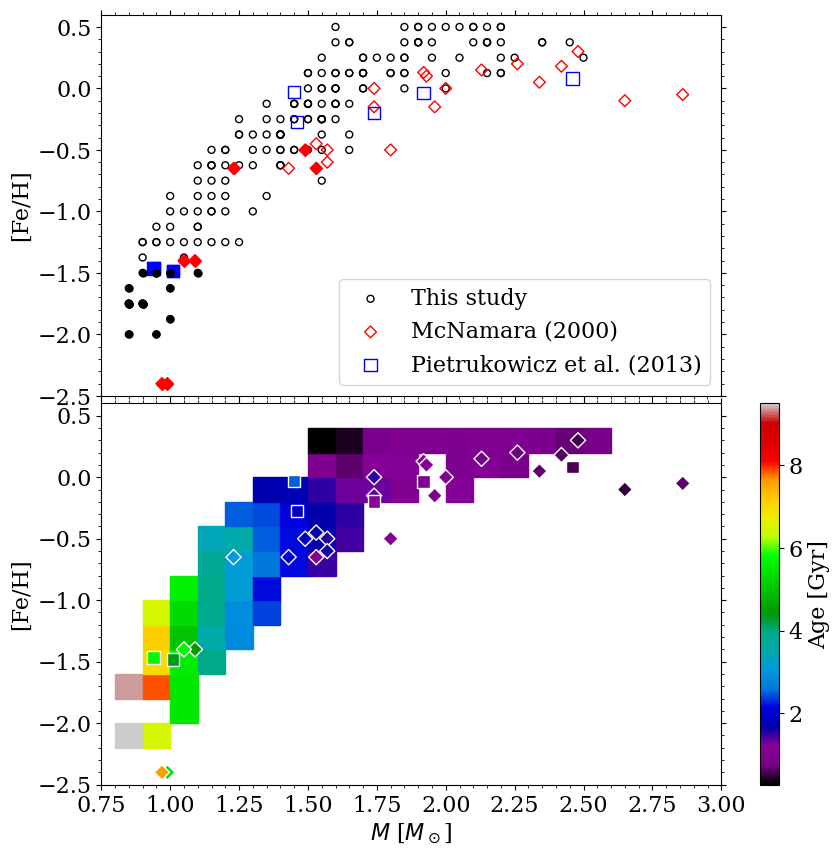}
\caption{Top panel: relation between mass and metallicity for HADS from this study, HADS from \protect\cite{pietrukowicz2013} and HADS from \protect\cite{mcnamara2000}. Stars classified as SX Phe are plotted with filled symbols. Bottom panel: age distribution for HADS stars. Stars from the sample of \protect\cite{pietrukowicz2013} and \protect\cite{mcnamara2000} are plotted with the same symbols as in the top panel and colors correspond to age as indicated in the key. Stars from this sample are shown as the distribution of mean ages in the bins for better visualisation (some stars overlap).}
\label{Fig.comparison_feh}
\end{figure}

In Fig.~\ref{Fig.comparison_feh} we compare masses and metallicities determined for HADS from this study to HADS studied by \cite{pietrukowicz2013} and by \cite{mcnamara2000}. In the top panel, we plotted $\delta$~Scuti stars with open symbols and stars classified as SX Phe in all three samples with filled symbols. The ranges of parameters covered by these three samples slightly differ. \cite{mcnamara2000} sample includes two stars with masses higher than 2.5\,M$_\odot$. Such high masses were not included in our grid. Except for these high-mass stars, the rest are within mass range of $0.8-2.5$\,M$_\odot$, with most stars within mass range of $1.5-2.0$\,M$_\odot$, similarly to our results. Also HADS from \cite{pietrukowicz2013} cover the same range as stars in this study. 

\cite{mcnamara2000} also reported two stars with very low metallicity (both have [Fe/H]$=-2.40$), which is also a value not included in our grid. These two stars from the sample by \cite{mcnamara2000} together with two stars of metallicities around $1.5$ and with three stars of metallicities around $-0.5$ are classified in the SIMBDAD database as SX~Phe.  HADS from \cite{pietrukowicz2013} are split into two groups with either close to solar metallicity or with metallicity around $-1.5$ (classified as SX~Phe). In our sample we observe a continuous group extending to low metallicities. With our criterion, only the lowest metallicity stars were classified as SX~Phe.

In the bottom panel of Fig.~\ref{Fig.comparison_feh} we plotted stars from the same samples as in the top panel of Fig.~\ref{Fig.comparison_feh}, but with colors we indicated age. Stars from this sample are plotted as distribution of mean ages in the bins for better visualisation, as some of the models overlap. For all three samples the age decreases with increasing masses and increasing metallicity. Above the mass of 1.5~M$_\odot$ the ages are below 2 Gyr. The highest correspond to stars on the low-mass and low-metallicity tail of the distribution.

\section{Conclusions}

In the presented study we used stars classified by \cite{netzel_hads} as multimode radial $\delta$~Scuti stars to obtain their physical parameters through asteroseismic modeling. We used stars that pulsate in at least three radial modes simultaneously. The input sample consists of 428 stars (145 F+1O+2O, 221 F+1O+3O, 48 1O+2O+3O and 14 F+1O+2O+3O). The method of modeling was based on the pre-calculated grid of models, where the physical parameters are determined by matching observed periods and period ratios with the calculated periods and period ratios for models in the grid. We did not use the additional observational constraints based e.g., on spectroscopy, since such observations are not available for the stars from the studied sample. The method used is therefore not a dedicated modeling of individual stars, but it allows us to study the statistical properties of $\delta$~Scuti sample as a whole. 

We calculated the grid of models using MESA evolution code \citep{mesa1,mesa2,mesa3,mesa4,mesa5} and Warsaw pulsation code \citep{dziembowski1977}.  This grid covers masses from 0.8\,M$_{\odot}$ to 2.5\,M$_{\odot}$, and metallicities [Fe/H] from $-2.0$ to $+0.5$. We used exponential scheme for convective overshooting from the convective core \citep{herwig2000}. We set the overshooting parameter, $f$, to 0.0 (no overshooting), 0.01, 0.02 or 0.03. Then we searched for the best fitting model for each star from the input sample by minimizing the difference between periods and periods ratios between models and observations. We obtained satisfactory fits for 176 HADS (2 F+1O+2O+3O stars, 68 F+1O+2O stars, 81 F+1O+3O stars and 25 1O+2O+3O stars) and reported masses, metallicities, ages, luminosities and effective temperatures for these stars. We classified stars based on the central hydrogen abundance as main sequence stars or post main sequence stars. We also classified stars as SX Phe based on their masses ($M\leqslant$ 1.1 M$_\odot$), ages (age $\geqslant 3$ Gyr ) and metallicities ([Fe/H] $\leqslant -1.5$).  The most important results are listed as follows:

\begin{itemize}
\item Stars pulsating in F+1O+2O, F+1O+3O and F+1O+2O+3O cover similar ranges in luminosity and effective temperature. Stars pulsating in 1O+2O+3O have, on average, higher luminosity.
\item Majority of stars pulsating in F+1O+2O, F+1O+3O and F+1O+2O+3O have masses within a range $0.8-2.0$\,M$_\odot$. On average, stars pulsating in 1O+2O+3O have higher masses, above 2\,M$_\odot$, as expected from their higher luminosities.
\item Typically, metallicity of all HADS is near solar from ${\rm [Fe/H]}=-0.5$ to ${\rm [Fe/H]}=+0.5$. There is a low-metallicity tail as well. Stars pulsating in 1O+2O+3O are the most numerous among the high-metallicity stars.
\item An average age in the sample is 1.7\,Gyr and is similar when considering all types of pulsations. The highest age among the sample is 9.5\,Gyr.
\item The adopted method of the modeling did not allow to put constraints on the extent of overshooting from the convective core.
\item We did not detect any clear relation between location of HADS in the Galaxy and their physical parameters.
\item 56 per cent of HADS are at the main sequence stage of evolution. 44 per cent already evolved of the main sequence.
\item There is a clear correlation of increasing metallicity of the fitted models with increasing mass. Increase in mass and metallicity is also correlated with lower ages of the models.
\item We selected 16 candidates for SX~Phe stars based on their higher ages, lower masses and lower metallicites.
\end{itemize}

The presented work is an excellent first step to study individual objects with precise asteroseismic modeling, when additional observational constraints, such as spectroscopic measurements of metallicity, effective gravity or effective temperature, become available. The follow-up work could include interpolation in a grid with more dense spacing in the overshooting parameter, $f_{\rm ov}$, with the hope to constrain the extent of convective core overshoot. Additional observational constraints should also enable to extend the modelling parameter space, e.g. with the mixing-length, $\alpha_{\rm MLT}$, which in this work was set to a constant, solar-calibrated value.

\section*{Acknowledgements}

This project has been supported by the Lend\"ulet Program of the Hungarian Academy of Sciences, project No. LP2018-7/2020. HN is supported in by the Polish Ministry of Science and Higher Education under grant 0192/DIA/2016/45 within the Diamond Grant Programme for years 2016-2021 and Foundation for Polish Science (FNP).  RS is supported by the National Science Center, Poland, Sonata BIS project 2018/30/E/ST9/00598.

This work has made use of data from the European Space Agency (ESA) mission
{\it Gaia} \url {https://www.cosmos.esa.int/gaia}), processed by the {\it Gaia}
Data Processing and Analysis Consortium (DPAC,
\url{https://www.cosmos.esa.int/web/gaia/dpac/consortium}). Funding for the DPAC
has been provided by national institutions, in particular the institutions
participating in the {\it Gaia} Multilateral Agreement.

\section*{Data Availability}

Photometric data used for this study were published as a part of
the OGLE project \citep{pietrukowicz2020} and are available at
the OGLE online data archive ({\UrlFont http://www.astrouw.edu.pl/ogle/ogle
4/OCVS/blg/dsct/}). The grid of evolutionary models was calculated with publicly available software MESA \citep{mesa5} and the information about the parameters is provided in the publication. The grid of pulsation models will be shared upon a reasonable request to the corresponding author.



\bibliographystyle{mnras}
\bibliography{references} 




\appendix
\newpage
\section{MESA inlist}\label{inlist}

\begin{verbatim}
&binary_controls

/ ! end of binary_controls namelist

&star_job

  initial_zfracs = 6
  pgstar_flag = .false.
  create_pre_main_sequence_model = .false.
  load_saved_model = .false.
  relax_z=.true.
  new_z=0.0004
  relax_y=.true.
  new_y=0.2701

  pause_before_terminate = .false.
 
  change_initial_net = .true. 
  new_net_name = 'pp_and_cno_extras.net'

  set_rate_c12ag = 'Kunz'
  set_rate_n14pg = 'jina reaclib'

  kappa_file_prefix = 'a09'
  kappa_lowT_prefix= 'lowT_fa05_a09p'
  kappa_CO_prefix = 'a09_co'

/ ! end of star_job namelist

&controls

  x_logical_ctrl(1) = .true.

  use_gold_tolerances  = .true.
  use_dedt_form_of_energy_eqn = .true.

  initial_mass = 2.0
  MLT_option = 'Cox'                  
  mixing_length_alpha = 1.76
        
  varcontrol_target = 1d-5
  mesh_delta_coeff = 0.5
  delta_HR_limit = 0.002

  use_Type2_opacities = .false. 
  Zbase = -1                    

! type of opacity interpolation in X/Z
  cubic_interpolation_in_X = .true.
  cubic_interpolation_in_Z = .true.

  which_atm_option = 'photosphere_tables'

  Teff_lower_limit  = 5500

  alpha_semiconvection = 0.1


  ! Mixing

  num_cells_for_smooth_gradL_composition_term = 20
  threshold_for_smooth_gradL_composition_term = 0.02

  num_cells_for_smooth_brunt_B = 50
  threshold_for_smooth_brunt_B = 1E-4

  use_ledoux_criterion = .true.

  do_conv_premix = .true.

  recalc_mix_info_after_evolve = .true.

  smooth_convective_bdy = .false.

   
  ! Overshoot
      overshoot_f_below_nonburn_shell = 0.015
      overshoot_f_above_burn_h_core = 0.0

      overshoot_f0_below_nonburn_shell = 0.005
      overshoot_f0_above_burn_h_core = 0.0


      photo_interval = 1000
      profile_interval = 1
      history_interval = 1
      terminal_interval = 50
      write_header_frequency = 50

/ ! end of controls namelist

&pgstar

/ ! end of pgstar namelist
\end{verbatim}




\bsp	
\label{lastpage}
\end{document}